\documentclass[12pt,a4paper]{article}

\newcommand{\la}{\lambda}
\newcommand{\ga}{\tilde{\gamma}}
\newcommand{\pa}{\partial}
\newcommand{\mj}{\mathcal{J}}
\newcommand{\tla}{\tilde{\lambda}}

\begin{document}

\begin{flushright}
{}
\end{flushright}
\vspace{1.8cm}

\begin{center}
 \textbf{\Large Rotating Strings with Two Unequal Spins \\
 in Lunin-Maldacena Background  }
\end{center}
\vspace{1.6cm}
\begin{center}
 Shijong Ryang
\end{center}

\begin{center}
\textit{Department of Physics \\ Kyoto Prefectural University of Medicine
\\ Taishogun, Kyoto 603-8334 Japan}  \par
\texttt{ryang@koto.kpu-m.ac.jp}
\end{center}
\vspace{2.8cm}
\begin{abstract}
We study a string motion in the Lunin-Maldacena background, that is,
the $\beta$-deformed $AdS_5 \times \tilde{S}^5$ background dual to
a $\beta$-deformation of $\mathcal{N} = 4$ super Yang-Mills theory.
For real $\beta$ we construct a rotating and wound string solution
which has two unequal spins in $\tilde{S}^5$. The string energy is
expressed in terms of the spins, the winding numbers and the deformation
parameter. In the expansion of $\la/J^2$ with the total spin $J$
and the string tension $\sqrt{\la}$ we present ``one-loop" and
``two-loop" energy corrections. The ``one-loop" one agrees with the 
one-loop anomalous dimension of the corresponding gauge-theory scalar
operators obtained in hep-th/0503192 from the $\beta$-deformed Bethe 
equation as well as the anisotropic Landau-Lifshitz equation.
\end{abstract} 
\vspace{3cm}
\begin{flushleft}
September, 2005
\end{flushleft}

\newpage
\section{Introduction}

The AdS/CFT correspondence \cite{MGW} has been explored beyond the 
supergravity approximation \cite{BMN,GKP}. In the BMN limit \cite{BMN}
for the $\mathcal{N} = 4$ super Yang-Mills (SYM) theory the perturbative
scaling dimensions of gauge invariant near-BPS operators with large
R-charge can be matched with the energies of certain string states, which
has been interpreted as the semiclassical quantization of nearly 
point-like string with large angular momentum along the central circle
of $S^5$ \cite{GKP}. Moreover, the energies of various semiclassical 
string with several large angular momenta in $AdS_5 \times S^5$ have been
shown in \cite{FRM,FT,BMSZ,AFRT,ART,AS,AT,SS} to match with the
anomalous dimensions of the corresponding gauge invariant non-BPS
operators which can be computed by using the Bethe ansatz \cite{MZ}
for diagonalization of the dilatation operator 
\cite{BKS,NB,NBS,BDS,NBP}, that is represented by 
a Hamiltonian of an integrable spin chain. 
From the view point of integrability the gauge/string
duality has been further confirmed by verifying the equivalence between
the classical string Bethe equation for the classical $AdS_5 \times S^5$
string sigma model and the Bethe equation for the spin chain in 
the various sectors such as SU(2), SL(2), SO(6) and so \cite{KMMZ,KZ},
while it has been demonstrated at the level of effective action where 
an interpolating spin chain sigma model action describing the continuum
limit of the spin chain in the coherent basis was constructed 
\cite{MK,KRT,AAT}.

In \cite{LM} Lunin and Maldacena have found a supergravity background
dual to the Leigh-Strassler or $\beta$-deformation of $\mathcal{N} = 4$ 
SYM theory \cite{LS} by applying a sequence of T-duality transformations
and shifts of angular coordinates to the original $AdS_5 \times S^5$
background. They have taken the plane-wave limit of the deformed
$AdS_5 \times \tilde{S}^5$ background with 
a deformed five-sphere $\tilde{S}^5$
and have shown that the string spectrum in the pp-wave coincides with 
the spectrum of BMN-type operators in the $\beta$-deformed 
$\mathcal{N} = 4$ SYM \cite{NP}. 

The Lax representation for the bosonic string theory in the real 
$\beta$-deformed background has been constructed \cite{SF}, which is 
similar to the undeformed case \cite{AF}. The gauge/string duality 
for the $\beta$-deformed case has been 
investigated by comparing the energies
of semiclassical strings to the anomalous dimensions of the gauge theory
operators in the two-scalar sector \cite{FRT}. The energy of circular
string with two equal angular momenta in $\tilde{S}^5$ has been computed
from the semiclassical approach and has been also derived from the 
anisotropic Landau-Lifshitz equation for the interpolating string sigma
model action which was obtained by taking 
the ``fast-string" limit of the world
sheet action in $AdS_5 \times \tilde{S}^5$. This interpolating action on
the string theory side has been shown to coincide with the continuum limit
of the coherent state action of an anisotropic XXZ spin chain on the 
$\beta$-deformed $\mathcal{N} = 4$ SYM theory side, 
which was constructed using the 
one-loop dilatation operator in the deformed gauge theory \cite{RR,BC}.
The classical string Bethe equation on the string theory side has been 
derived from the Lax representation of \cite{SF} to coincide with the
thermodynamic limit of the Bethe equation for the anisotropic spin chain.
Various relevant aspects of the gauge/string duality for the 
marginal deformed backgrounds have been investigated 
\cite{KMSS,BDR,BR,FG,GN}.

Multi-spin configurations of strings which move in both $AdS_5$ 
and $\tilde{S}^5$ parts of the $\beta$-deformed background have been
constructed \cite{BDR}, where the winding numbers and the frequencies
associated with the angular momenta take the same magnitudes 
respectively for each part of $AdS_5 \times \tilde{S}^5$.
This property was also seen in the circular string solution with
two equal spins \cite{FRT} which is specified by
 the same frequencies and winding numbers.
We will construct a circular string solution with two unequal spins in
$\tilde{S}^5$ which is characterized by different frequencies and
winding numbers. The energy of the string solution will be computed
and represented in terms of the unequal winding numbers, the unequal
angular momenta and the deformation parameter.
This energy spectrum on the string theory side will be compared with
that of the solution in \cite{FRT} for the anisotropic Landau-Lifshitz
equation and the $\beta$-deformed Bethe equation for the spin 
configuration with the filling fraction away from one half.

\section{Rotating string solution with two unequal spins}

We consider a rotating closed string motion in the supergravity 
background dual to the real $\beta$-deformation of the $\mathcal{N}=4$
SYM theory. The $\beta$-deformed background for real $\beta = \gamma$
is given by \cite{LM}
\begin{eqnarray}
ds_{str}^2 &=& R^2 \left[ ds_{AdS_5}^2 + \sum_{i=1}^{3}(d\rho_i^2 +
G\rho_i^2 d\phi_i^2 ) +  \ga^2G \rho_1^2 \rho_2^2 \rho_3^2(\sum_{i=1}^{3}
d\phi_i)^2 \right], \nonumber \\
B_2 &=& R^2 \ga G w_2, \hspace{1cm} w_2 \equiv 
\rho_1^2\rho_2^2d\phi_1d\phi_2   + \rho_2^2\rho_3^2d\phi_2d\phi_3 + 
\rho_3^2\rho_1^2d\phi_3d\phi_1,   \\
G^{-1} &=& 1 + \ga^2 Q, \hspace{1cm} Q \equiv \rho_1^2\rho_2^2  + 
\rho_2^2\rho_3^2 + \rho_3^2\rho_1^2,  \hspace{1cm} \sum_{i=1}^{3}\rho_i^2
 = 1, \nonumber 
\end{eqnarray}
where 
\begin{equation}
(\rho_1, \rho_2, \rho_3) = (\sin\alpha \cos \theta, \sin \alpha 
\sin\theta, \cos\alpha), \hspace{1cm} R^4 = Ng_{YM}^2 = \la 
\end{equation}
and the regular
deformation parameter $\ga$ of the supergravity background is related with
the real deformation parameter $\gamma$ of the deformed $\mathcal{N}=4$
SYM as $\ga = R^2 \gamma$. We concentrate on a configuration that a closed
string is staying at the center of $AdS_5$ and moving in the 
$\tilde{S}^3$ part of the deformed five-sphere defined by
\begin{equation}
\alpha = \frac{\pi}{2}, \hspace{1cm} \mathrm{i.e.}  \hspace{1cm}
\rho_1 = \cos\theta, \; \rho_2 = \sin\theta, \; \rho_3 = 0. 
\end{equation}
The relevant bosonic string action takes the form
\begin{eqnarray}
S &=& -\frac{1}{2} R^2\int d\tau \int \frac{d\sigma}{2\pi} \Bigl[
\gamma^{\alpha\beta}(-\pa_{\alpha}t \pa_{\beta}t + \pa_{\alpha}\theta 
\pa_{\beta}\theta + G\cos^2\theta \pa_{\alpha}\phi_1 \pa_{\beta}\phi_1
+ G\sin^2\theta \pa_{\alpha}\phi_2 \pa_{\beta}\phi_2 ) \nonumber \\
 & & - 2\epsilon^{\alpha\beta} \ga G \sin^2\theta \cos^2\theta 
\pa_{\alpha}\phi_1 \pa_{\beta}\phi_2 \Bigr], 
\end{eqnarray}
where $\epsilon^{01}=1, \gamma^{\alpha\beta}$ is expressed as
$\gamma^{\alpha\beta}=\sqrt{-h} h^{\alpha\beta}$ in terms of a world-sheet
metric $h^{\alpha\beta}$, and 
\begin{equation}
G = \frac{1}{1 + \frac{\ga^2}{4}\sin^22\theta}.
\end{equation}

We choose the conformal gauge $\gamma^{\alpha\beta}= \mathrm{diag}(-1,1)$
and make the following ansatz describing a closed string rotating and
wound in the $\phi_1$ and $\phi_2$ directions 
\begin{equation}
t = \kappa\tau, \hspace{1cm} \phi_1 = \omega_1 \tau + m_1\sigma, 
\hspace{1cm} \phi_2 = \omega_2 \tau + m_2\sigma, \hspace{1cm} 
\theta = \theta_0 = \mathrm{const},
\end{equation}
where $m_1, m_2$ are the winding numbers. The string equation of motion
for $\theta$ is satisfied when the constant $\theta_0$ is specified by
\begin{eqnarray}
\left[\omega_1^2 - m_1^2 - (\omega_2^2 - m_2^2)\right] \left( 1 + 
\frac{\ga^2}{4}\sin^22\theta_0 \right) -2\ga 
(\omega_1 m_2 - \omega_2m_1)\cos 2\theta_0 \nonumber \\
+ \ga^2 \cos2\theta_0 \left[\cos^2\theta_0(\omega_1^2 - m_1^2)
+ \sin^2\theta_0(\omega_2^2 - m_2^2) \right] = 0.
\label{thc}\end{eqnarray}
It has a simple string solution with two equal angular momenta, which is
described by $\theta_0 = \pi/4, \; \omega_1^2 - m_1^2 = \omega_2^2 - 
m_2^2$ in ref. \cite{FRT}. Here we look for an extended solution with 
two unequal angular momenta which reduces to the simple solution in
a special parameter limit. The equation (\ref{thc}) is rewritten in terms
of $x = \cos2\theta_0$ as 
\begin{equation}
\frac{\ga^2}{4}(\Omega_1 - \Omega_2)x^2 + \ga\left( \frac{\Omega_1 + 
\Omega_2}{2} \ga - 2\Omega_0  \right) x + 
\left(1 + \frac{\ga^2}{4} \right)(\Omega_1 - \Omega_2) = 0,
\label{omx}\end{equation}
where $\Omega_i \equiv \omega_i^2 - m_i^2 \; (i=1, 2), \;
\Omega_0 \equiv \omega_1m_2 - \omega_2m_1$. This equation
determines $x$ in terms of $\omega_i, \; m_i(i=1,2)$ and $\ga$.
The angular momenta $\mj_1 = J_1/\sqrt{\la}$ and $\mj_2 = J_2/\sqrt{\la}$
coming from the rotations in the $\phi_1$ and $\phi_2$ directions
are obtained by
\begin{eqnarray}
\mj_1 &=& \frac{1}{1 + \frac{\ga^2}{4}(1-x^2)}\left[ \frac{1+x}{2}\omega_1
+ \frac{\ga}{4}( 1 - x^2 )m_2 \right], \label{tao} \\
\mj_2 &=& \frac{1}{1 + \frac{\ga^2}{4}(1-x^2)}\left[ \frac{1-x}{2}\omega_2
- \frac{\ga}{4}( 1 - x^2 )m_1 \right].
\label{tat}\end{eqnarray}
From them the frequencies $\omega_1$ and $\omega_2$ are expressed as 
\begin{eqnarray}
\omega_1 &=& \frac{2}{1+x}\left[ \left( 1 + \frac{\ga^2}{4}(1-x^2) \right)
\mj_1 - \frac{\ga}{4}( 1 - x^2 )m_2 \right],  \label{omo} \\
\omega_2 &=& \frac{2}{1-x}\left[ \left( 1 + \frac{\ga^2}{4}(1-x^2) \right)
\mj_2 + \frac{\ga}{4}( 1 - x^2 )m_1 \right].
\label{omt}\end{eqnarray}

The conformal gauge constraints imply
\begin{eqnarray}
\kappa^2 = \frac{1}{1 + \frac{\ga^2}{4}(1-x^2)} \left[ \frac{1 + x}{2}
(\omega_1^2 + m_1^2 )  + \frac{1 - x}{2}(\omega_2^2 + m_2^2 ) 
\right], \label{kax} \\
\frac{1 + x}{2}\omega_1m_1 + \frac{1 - x}{2}\omega_2m_2 = 0.
\label{omm}\end{eqnarray}
The substitution of $\omega_1$ and $\omega_2$ in (\ref{omo}) and 
(\ref{omt}) into the eq. (\ref{omm}) leads to a compact expression
\begin{equation}
\mj_1 m_1 + \mj_2 m_2 = 0.
\label{tam}\end{equation}
It is noted that this expression for the $\gamma$-deformed background 
takes the same form as that for the undeformed background. 
When $\theta_0 = \pi/4$, that is, $x = 0$, the eqs. (\ref{tao}), 
(\ref{tat}) and (\ref{omm}) provide $\mj_1 = \mj_2 = \mj/2, \; \omega_1 
= \omega_2 = \mj + \ga(m + \ga\mj/2)/2$ with $m_1 = -m_2 \equiv m$
and the total spin $\mj = \mj_1 + \mj_2$, which
is the special string solution with two equal angular momenta \cite{FRT}.
From (\ref{kax}) the energy of circular string solution is specified by
\begin{equation}
E^2 = \frac{\la}{1 + \frac{\ga^2}{4}(1-x^2)} \left[ \frac{1 + x}{2}
(\omega_1^2 + m_1^2 )  + \frac{1 - x}{2}(\omega_2^2 + m_2^2 ) \right].
\label{exo}\end{equation}
The quadratic equation (\ref{omx}) yields a solution expressed in terms of
$\omega_1, \omega_2$, which is further inserted into (\ref{tao}), 
(\ref{tat}). If we can determine  $\omega_1, \omega_2$ as functions of
$\mj_1, \mj_2$ from the two inserted equations, we substitute these
functions into (\ref{exo}) to obtain the energy expressed in terms of
$\mj_1, \mj_2$. However, it is impossible to 
derive the functions so that we take
the following alternative procedure. First combining (\ref{omo}) and
(\ref{omt}) with (\ref{exo}) we have the energy expression
\begin{eqnarray}
E^2 &=& \la \Biggl[ 2\left( 1 + \frac{\ga^2}{4}(1-x^2) \right) 
\left( \frac{\mj_1^2}{1+x} + \frac{\mj_2^2}{1-x} \right) + \ga( - m_2\mj_1
+ m_1\mj_2 ) \nonumber \\
&+& \ga( m_2\mj_1 + m_1\mj_2 )x + \frac{1 + x}{2}m_1^2 + 
\frac{1 - x}{2}m_2^2 \Biggr].
\label{ext}\end{eqnarray}
When $\omega_1$ and $\omega_2$ in (\ref{omo}) and (\ref{omt}) are directly
substituted into (\ref{omx}) we have an involved equation for $x$.
If a solution $x$ of the transcendental equation is obtained as a function
of $\mj_i, m_i (i=1,2), \ga$ and inserted into (\ref{ext}), then
the energy of string solution is expressed in terms of the angular 
momenta, the winding numbers and the deformation parameter.

\section{Energy-spin relation}

In order to solve the transcendental equation
we consider the parameter region specified by $x \ll 1$ and take the 
expansion around $x = 0$. This is the case of almost equal spins, i.e.
$\Delta J \ll J, \; \Delta J \equiv J_1 - J_2$. We use (\ref{omo})
and (\ref{omt}) to expand $\Omega_1 - \Omega_2$ in (\ref{omx})
in powers of $x$ as 
\begin{equation}
\Omega_1 - \Omega_2 = A_0 + A_1x + A_2x^2 + A_3x^3 + \cdots ,
\label{oma}\end{equation}
where
\begin{eqnarray}
&A_0 = 4\left(1 + \frac{\ga^2}{4}\right) \left[ \left(1 + 
\frac{\ga^2}{4}\right)(\mj_1^2 - \mj_2^2) - \frac{\ga}{2}
(m_2\mj_1 + m_1\mj_2) - \frac{1}{4} (m_1^2 - m_2^2) \right],& \nonumber \\
&A_1 = -8 \left[\left(1 + \frac{\ga^2}{4}\right)^2(\mj_1^2 + \mj_2^2)
+ \frac{\ga}{2}\left(1 + \frac{\ga^2}{4}\right)( - m_2\mj_1
+ m_1\mj_2 ) + \frac{\ga^2}{16} (m_1^2 + m_2^2) \right],& \nonumber \\
&A_2 = 12\left(1 + \frac{\ga^2}{4}\right)\left(1 + \frac{\ga^2}{12}
\right) (\mj_1^2 - \mj_2^2) - 4\ga\left(1 + \frac{\ga^2}{8}\right)
(m_2\mj_1 + m_1\mj_2) - \frac{\ga^2}{4} (m_1^2 - m_2^2),&  \nonumber \\
&A_3 = -16 \left[\left(1 + \frac{\ga^2}{4}\right)(\mj_1^2 + \mj_2^2)
+ \frac{\ga}{4}( - m_2\mj_1 + m_1\mj_2 ) \right]. &
\end{eqnarray}
Similarly, $(\Omega_1 + \Omega_2)\ga/2 - 2\Omega_0$ in (\ref{omx})
can be expanded as 
\begin{equation}
\frac{\Omega_1 + \Omega_2}{2} \ga - 2\Omega_0  = B_0 + B_1x + B_2x^2 + 
\cdots, \nonumber
\label{omb}\end{equation}
where 
\begin{eqnarray}
&B_0 = 2\left(1 + \frac{\ga^2}{4}\right)^2 \left[\ga(\mj_1^2 + \mj_2^2)
+ 2( - m_2\mj_1 + m_1\mj_2 ) + \frac{\ga}{4} \frac{m_1^2 + m_2^2}
{1 + \frac{\ga^2}{4}} \right], &  \nonumber \\
&B_1 = -4\left(1 + \frac{\ga^2}{4}\right) \left[ \ga\left(1 + 
\frac{\ga^2}{4}\right)(\mj_1^2 - \mj_2^2) - \left( 1 + \frac{\ga^2}{2}
\right)(m_2\mj_1 + m_1\mj_2) - \frac{\ga}{4} 
(m_1^2 - m_2^2) \right],&  \nonumber \\
&B_2 = 6\left(1 + \frac{\ga^2}{4}\right) \left[\ga\left(1 + 
\frac{\ga^2}{12}\right)(\mj_1^2 + \mj_2^2) + \frac{2}{3}\left(1 + 
\frac{\ga^2}{4}\right) ( - m_2\mj_1 + m_1\mj_2 ) 
  + \frac{\ga^3}{48} 
\frac{m_1^2 + m_2^2}{1 + \frac{\ga^2}{4}} \right]. &
\end{eqnarray}
It is noted that the coefficients $A_k, B_k \;(k=0,1,2,\cdots)$ show the
alternate expressions for $k=$ even and $k=$ odd.
Hence by combining (\ref{oma}) and (\ref{omb}) with (\ref{omx}) and
taking account of the behaviors that $A_0$ is of order $\epsilon$, while
$A_1$ and $B_0$ are of order $\epsilon^0$, we
make the leading order estimation of $x$ as 
\begin{equation}
x_1 = - \frac{\left(1 + \frac{\ga^2}{4}\right)A_0}{\ga B_0
+ \left(1 + \frac{\ga^2}{4}\right)A_1 },
\end{equation}
which is rewritten by
\begin{equation}
x_1 = \frac{1 + \frac{\ga^2}{4}}{2(\mj_1^2 + \mj_2^2)}
\left[ \mj_1^2 - \mj_2^2 - \frac{\ga}{2} \frac{m_2\mj_1 + m_1\mj_2}
{1 + \frac{\ga^2}{4}} - \frac{m_1^2 - m_2^2}{4\left(1 + 
\frac{\ga^2}{4}\right) } \right].
\label{xmn}\end{equation}
Using the Virasoro constraint (\ref{tam}) and 
a parameter $\alpha = \mj_1/\mj$ with the total
spin $\mj$ we illustrate that $x_1$ is the term of order
$\epsilon = 2\alpha - 1$ 
\begin{equation}
x_1 = \frac{2\alpha - 1}{1 + (2\alpha - 1)^2} \left[ 1 + \frac{\ga^2}{4}
+ \frac{\ga}{2} \frac{m_1}{\mj_2} + \frac{1}{4}\left( \frac{m_1}{\mj_2} 
\right)^2 \right].
\end{equation}

It is possible to estimate the non-leading term $x_2$ by substituting
$x = x_1 + x_2$ into the eq. (\ref{omx}) accompanied with (\ref{oma})
and (\ref{omb}). By considering the behaviors that $A_0, A_2$ and
$B_1$ are of order $\epsilon$, while $A_1, A_3, B_0$ and $B_2$
are of order $\epsilon^0$, we obtain
\begin{eqnarray}
x_2 = -\frac{1}{\ga B_0 + \left(1 + \frac{\ga^2}{4}\right)A_1 }
\Biggl[ x_1^2 \left(\frac{\ga^2}{4}A_0 + \ga B_1 + \left(1 + 
\frac{\ga^2}{4}\right)A_2 \right) \nonumber \\
+ x_1^3\left( \frac{\ga^2}{4}A_1 + \ga B_2 + \left(1 + 
\frac{\ga^2}{4}\right)A_3 \right) \Biggr],
\end{eqnarray}
which is the term of order $\epsilon^3$ as shown by
\begin{eqnarray}
x_2 &=& \frac{x_1^2}{\mj_1^2 + \mj_2^2} \left[ \frac{3}{2}
\left(1-\frac{\ga^2}{6}\right)(\mj_1^2 - \mj_2^2) + \frac{\ga^3}{8}
\frac{m_2\mj_1 + m_1\mj_2}{1 + \frac{\ga^2}{4}} + \frac{\ga^2}{16}
\frac{m_1^2 - m_2^2}{1 + \frac{\ga^2}{4}} \right] \nonumber \\
 &-& \frac{2x_1^3}{1 + \frac{\ga^2}{4}}.
\label{xtw}\end{eqnarray}
The substitution of $x = x_1 + x_2 + x_3$ where $x_3$ is the term 
of order $\epsilon^5$ into the energy expression (\ref{ext})
yields the following expansion up to order $\epsilon^6$
\begin{eqnarray}
E^2 &=& \la \Biggl[ 2(\mj_1^2 + \mj_2^2) + \frac{1}{2}(m_2 - \ga\mj_1)^2
+ \frac{1}{2}(m_1 + \ga\mj_2)^2 \nonumber \\
&+& (x_1 + x_2 + x_3)\left( -2\left(1 + \frac{\ga^2}{4}\right)
(\mj_1^2 - \mj_2^2) + \ga(m_2\mj_1 + m_1\mj_2) + 
\frac{m_1^2 - m_2^2}{2} \right) \nonumber \\
&+& 2(\mj_1^2 + \mj_2^2)(x_1^2 + 2x_1x_2 + 2x_1x_3 + x_2^2 + x_1^4 + 
4x_1^3x_2 + x_1^6 ) \\
&-& 2(\mj_1^2 - \mj_2^2)( x_1^3 + 3 x_1^2x_2 + x_1^5 ) + \cdots \Biggr].
\nonumber
\end{eqnarray}
Since the fourth term can be expressed as $(x_1 + x_2 + x_3)
(-4(\mj_1^2 + \mj_2^2)x_1)$, we have
\begin{eqnarray}
E^2 &=& P + 2(J_1^2 + J_2^2)(x_1^4 + x_2^2 + 4x_1^3x_2 + x_1^6)
-2(J_1^2 - J_2^2)(x_1^3 + 3x_1^2x_2 + x_1^5 ) + \cdots, \\
P &\equiv& 2(J_1^2 + J_2^2)( 1- x_1^2 ) + \frac{\la}{2}
(m_2 - \gamma J_1)^2 + \frac{\la}{2}(m_1 + \gamma J_2)^2, \nonumber
\end{eqnarray}
where the $x_1x_3$ term has been canceled out. The expression $x_1$
in (\ref{xmn}) is rewritten by
\begin{equation}
x_1 = \frac{1}{1 + (2\alpha -1)^2} \left[ 2\alpha -1 + \frac{\tla}{4}
(m_2 - \gamma J_1)^2 - \frac{\tla}{4}(m_1 + \gamma J_2)^2 \right]
\equiv \frac{\bar{x}_1}{1+(2\alpha-1)^2}, 
\label{xon}\end{equation}
where $\tla$ is the effective coupling constant $\tla = \la/J^2 =
1/\mj^2$, so that $P$ takes the form
\begin{eqnarray}
P &=& P_0 - \frac{\tla^2J^2}{16}\left[(m_2 - \gamma J_1)^2 - 
(m_1 + \gamma J_2)^2\right]^2 + J^2(2\alpha -1)^2(1 + (2\alpha -1)^2)
x_1^2, \\
P_0 &\equiv& J^2\left[ 1 - \frac{\tla(2\alpha -1)}{2} \left(
(m_2 - \gamma J_1)^2 - (m_1 + \gamma J_2)^2 \right) \right]
+ \frac{\la}{2}(m_2 - \gamma J_1)^2 + \frac{\la}{2}
(m_1 + \gamma J_2)^2. \nonumber
\end{eqnarray}
The leading part $P_0$ which is of order $\epsilon^0$ is rearranged
into
\begin{equation}
P_0 = J^2\left[ 1 + \frac{J_1J_2}{J^2} \tla
(\gamma J + m_1 - m_2)^2 + \frac{\tla}{J^2}(m_1J_1 + m_2J_2 )^2 \right],
\end{equation}
which is expressed through the Virasoro constraint (\ref{tam}) as
\begin{equation}
P_0 = J^2 + J^2\alpha( 1 - \alpha )\tla(\gamma J + m_1 - m_2)^2.
\end{equation}
Gathering together with the expression $x_2$ in (\ref{xtw})
described by
\begin{eqnarray}
x_2 &=& \frac{x_1^2}{1 + (2\alpha -1)^2} \bar{x}_2,  \\
\bar{x}_2 &\equiv& \left(1-\frac{\ga^2}{2}
\right)(2\alpha -1) + \tla \gamma ( m_2J_1 + m_1J_2 )  + 
\frac{\tla}{2}(m_1^2 - m_2^2) \nonumber
\end{eqnarray}
we get the following energy expression 
\begin{eqnarray}
E^2 &=& P_0 - \frac{\tla^2J^2}{16}\left[(m_2 - \gamma J_1)^2 - 
(m_1 + \gamma J_2)^2\right]^2  \\
&+& \frac{\tla^2J^2}{16}\frac{x_1^2}{1 + (2\alpha -1)^2}\left[(m_2 - 
\gamma J_1)^2 - (m_1 + \gamma J_2)^2\right]^2 + 
2J^2\bar{x}_1^2(2\alpha -1)^4
+ P_6 + \cdots, \nonumber
\end{eqnarray}
where 
\begin{eqnarray}
P_6 &\equiv& J^2\bar{x}_1^4 \Biggl[ -2(2\alpha -1)\bar{x}_1  
+ (2\alpha-1)\tla(\gamma J)^2\bar{x}_2 + \bar{x}_1^2
+ \frac{\ga^4}{4}(2\alpha -1)^2 \nonumber \\
&-& \left( 2\alpha -1 + \tla \gamma ( m_2J_1 + m_1J_2 ) + 
\frac{\tla}{2}(m_1^2 - m_2^2) \right)^2 \Biggr].
\end{eqnarray}
The second term is of order $\epsilon^2$ and the third term is 
of order $\epsilon^4$ for the leading part, while the fourth term and
$P_6$ are of order $\epsilon^6$.

Now we use the resulting expression up to order $\epsilon^4$
as well as up to order $\tla^2$ 
\begin{equation}
E^2 = P_0 -  \frac{\tla^2J^2}{16}\left[(m_2 - \gamma J_1)^2 - 
(m_1 + \gamma J_2)^2\right]^2( 1 - (2\alpha -1)^2) + \cdots
\end{equation}
to extract the energy of the solution as
\begin{eqnarray}
E &=& J + \frac{\la}{2J}\alpha(1-\alpha)(\gamma J + m_1 - m_2)^2 
\label{eja}   \\
&-& \frac{\la^2}{8J^3}\alpha(1-\alpha)\left[ \left((m_2 - \gamma J_1)^2 -
(m_1 + \gamma J_2)^2 \right)^2 + \alpha(1-\alpha)
(\gamma J + m_1 - m_2)^4\right] + \cdots. \nonumber
\end{eqnarray}
The second term of order $\tla$, that is, the ``one-loop" energy
correction agrees with the one-loop result of ref. \cite{FRT}, where the
$\beta$-deformed Landau-Lifshitz action produced from 
``fast-string" expansion of the string sigma model action is shown to 
have a rational solution with two unequal momenta and the one-loop 
anomalous dimension is computed also by solving the Bethe equation for
the corresponding anisotropic spin chain. 

The undeformed limit $\gamma \rightarrow 0$ for (\ref{eja}) yields
\begin{equation}
E = J + \frac{\la}{2J}m(n-m) - \frac{\la^2}{8J^3}m(n-m)(n^2 -3nm + 3m^2)
+ \cdots,
\label{emn}\end{equation}
where the relations provided from (\ref{tam}) for the undeformed case
\begin{equation}
\frac{J_1}{J} = -\frac{m_2}{m_1 - m_2}, \hspace{1cm}  \frac{J_2}{J} = 
\frac{m_1}{m_1 - m_2}
\label{jra}\end{equation}
have been used and the winding numbers $m_1, m_2$ have been replaced
by $m_1 = n - m, \; m_2 = -m$. The reduced expression (\ref{emn}) 
including the ``one-loop" and ``two-loop" corrections was presented 
\cite{KMMZ} by analyzing the classical Bethe 
equation for the classical $AdS_5 \times
S^5$ string sigma model as well as the Bethe equation for the spin
chain in the SU(2) sector. On the other hand in ref. \cite{ART}
a general class of rotating string solutions in the undeformed
background was derived and the ``one-loop" correction in (\ref{emn})
was presented by solving  the relations for the two spin sector
\begin{equation}
\mathcal{E}^2 = 2\sum_{i=1}^2\omega_i\mj_i - \nu^2, \hspace{1cm}
\sum_{i=1}^2\frac{\mj_i}{\omega_i} = 1, \hspace{1cm}
\sum_{i=1}^2m_i\mj_i = 0,
\end{equation}
where $\omega_i^2 - m_i^2 = \nu^2\; (i=1,2)$ and the last equation indeed
takes the same form as (\ref{tam}). Here we demonstrate that these 
relations also yield the ``two-loop" correction. Indeed the
following large $\mj$ expansion
\begin{equation}
\mathcal{E}^2 = \mj^2 + \sum_{i=1}^2m_i^2\frac{\mj_i}{\mj} +
\frac{1}{4\mj^2}\left[ \left(\sum_{i=1}^2m_i^2\frac{\mj_i}{\mj}\right)^2
- \sum_{i=1}^2m_i^4\frac{\mj_i}{\mj} \right] + \cdots
\end{equation}
leads to 
\begin{equation}
E = \sqrt{\la}\mathcal{E} = J + \frac{\la}{2J}\sum_{i=1}^2m_i^2
\frac{J_i}{J} - \frac{\la^2}{8J^3}\sum_{i=1}^2m_i^4\frac{J_i}{J}  
+ \cdots.
\label{emj}\end{equation}
The substitution of $m_1 = n - m, \; m_2= -m$ and (\ref{jra}) into
(\ref{emj}) reproduces (\ref{emn}).

Following the argument of \cite{FRT}, the structure of the energy of 
a multi-spin solution can be captured as double expansions in $\tla$
and $\ga$. The smoothness of the deformation gives
\begin{equation}
E = \sqrt{\la}\mathcal{E}(\ga,\mj) = \sqrt{\la}\left[ \mathcal{E}_0(\mj) +
\ga f_1(\mj) + \ga^2 f_2(\mj) + \ga^3 f_3(\mj) + \ga^4 f_4(\mj) 
+ \cdots \right],
\label{ex}\end{equation}
while the energy for the undeformed case has the usual regular 
large $\mj$ or small $\tla$ expansion 
\begin{equation}
E_0 = \sqrt{\la}\mathcal{E}_0(\mj) = Jf_0(\tla) =
J(1 + c_1\tla + c_2\tla^2 + \cdots).
\end{equation}
Through $\ga = \sqrt{\tla}\gamma J$ the expansion (\ref{ex}) 
turns out to be
\begin{equation}
E = J\left[ f_0(\tla) + \tla \gamma Jf_1(\mj) + \tla (\gamma J)^2
\frac{f_2(\mj)}{\mj} + \tla^2 (\gamma J)^3f_3(\mj) +
\tla^2 (\gamma J)^4\frac{f_4(\mj)}{\mj} + \cdots \right].
\end{equation}
If $f_1(\mj), f_2(\mj)/\mj, f_3(\mj), f_4(\mj)/\mj$ have the
regular expansions in $1/\mj^2 = \tla$ as
\begin{equation}
f_1 = \sum_{k=0}^{\infty}c_k^1\tla^k, \hspace{1cm} \frac{f_2}{\mj} =  
\sum_{k=0}^{\infty}c_k^2\tla^k, \hspace{1cm} f_3 = \sum_{k=0}^{\infty}
c_k^3\tla^k, \hspace{1cm} \frac{f_4}{\mj} =  
\sum_{k=0}^{\infty}c_k^4\tla^k,
\end{equation}
then $E$ also takes the following regular expansion
\begin{eqnarray}
E &=& J\Bigl[ 1 + \tla\left(c_1 + c_0^1\gamma J + c_0^2(\gamma J)^2
\right)  \nonumber \\
 &+& \tla^2 \left( c_2 + c_1^1\gamma J + c_1^2(\gamma J)^2 + 
c_0^3(\gamma J)^3 + c_0^4(\gamma J)^4 \right) + \cdots \Bigr].
\end{eqnarray}
Hence we can see that the ``two-loop" part of order $\tla^2$ contains 
the terms $(\gamma J)^n$ with $n \le 4$. This behavior is indeed
seen in our explicit ``two-loop" result 
\begin{eqnarray}
E_2 &=&  -\frac{\la^2}{8J^3}\alpha(1-\alpha)\Bigl[ \left((1-2\alpha)\gamma
J  + m_1 + m_2\right)^2(\gamma J + m_1 - m_2)^2 \nonumber \\
&+& \alpha(1-\alpha)(\gamma J + m_1 - m_2)^4\Bigr].
\end{eqnarray}

\section{Conclusion}

Analyzing the closed string motion in the $\gamma$-deformed
$AdS_5 \times \tilde{S}^5$ background by the semiclassical
approach, we have presented a transcendental equation which determines
the energy of a solution describing a rotating and wound string
with two unequal spins in $\tilde{S}^5$. By using an expansion
procedure for the case of almost equal spins we have solved the
transcendental equation to extract the string energy in terms of the
angular momenta, the winding numbers and the deformation parameter.

The ``one-loop" part of the string energy expanded in $\la/J^2$
has reproduced the one-loop anomalous dimension of long two-scalar
operator which was found in ref. \cite{FRT} as a solution for the
anisotropic Landau-Lifshitz and Bethe equations for the 
$\gamma$-deformed $\mathcal{N}=4$ SYM
for the $SU(2)_{\gamma}$ sector. This agreement 
at the one-loop level confirms the gauge/string duality between the 
superstring theory in the $\gamma$-deformed $AdS_5 \times \tilde{S}^5$
background and the $\gamma$-deformed $\mathcal{N}=4$ SYM theory.
We have observed that the ``two-loop" part of the string energy
consists of the terms $(\gamma J)^n$ with $n \le 4$, and recovers
the corresponding expected ``two-loop" string energy for the undeformed
case when we take the undeformed limit $\gamma \rightarrow 0$.
In order to confirm the gauge/string duality at the two-loop level
it is desirable to construct some two-loop dilatation operator for 
the $\gamma$-deformed $\mathcal{N}=4$ SYM theory and compute the 
two-loop anomalous dimension of the relevant gauge invariant scalar
operator.


\begin{thebibliography}{99}
\bibitem{MGW} J.M. Maldacena, ``The large N limit of superconformal
field theories and supergravity,'' Adv. Theor. Math. Phys. \textbf{2}
(1998) 231 [hep-th/9711200]; S.S. Gubser, I.R. Klebanov and A.M. Polyakov,
``Gauge theory correlators from non-critical string theory,"
Phys. Lett. \textbf{B428} (1998) 105 [hep-th/9802109]; E. Witten, 
``Anti-de Sitter space and holography,"
Adv. Theor. Math. Phys. \textbf{2} (1998) 253 [hep-th/9802150].
\bibitem{BMN} D. Berenstein, J.M. Maldacena and H. Nastase, 
``Strings in flat space and pp waves from $\mathcal{N}$=4 super
Yang Mills," JHEP \textbf{04} (2002) 013 [hep-th/0202021].
\bibitem{GKP} S.S. Gubser, I.R. Klebanov and A.M. Polyakov,
``A semi-classical limit of the gauge/string correspondence,"
Nucl. Phys. \textbf{B636} (2002) 99 [hep-th/0204051].
\bibitem{FRM} S. Frolov and A.A. Tseytlin, ``Semiclassical 
quantization of rotating superstring in $AdS_5\times S^5$," JHEP
\textbf{06} (2002) 007 [hep-th/0204226];
J.G. Russo, ``Anomalous dimensions in gauge theories
from rotating strings in $AdS_5\times S^5$," JHEP \textbf{06} (2002)
038 [hep-th/0205244];
J.A. Minahan, ``Circular semiclassical string solutions
on $AdS_5\times S^5$," Nucl. Phys. \textbf{B648}
(2003) 203 [hep-th/0209047].
\bibitem{FT} S. Frolov and A.A. Tseytlin, ``Multi-spin string solutions
in $AdS_5 \times S^5$," Nucl. Phys. \textbf{B668} (2003) 77 
[hep-th/0304255]; ``Quantizing three-spin
string solution in $AdS_5 \times S^5$," JHEP \textbf{07} (2003) 016
[hep-th/0306130]; ``Rotating string 
solutions: AdS/CFT duality in non-supersymmetric sectors," Phys. Lett.
\textbf{B570} (2003) 96 [hep-th/0306143].
\bibitem{BMSZ} N. Beisert, J.A. Minahan, M. Staudacher and K. Zarembo,
``Stringing spins and spinning strings," JHEP \textbf{09}
(2003) 010 [hep-th/0306139];
N. Beisert, S. Frolov, M. Staudacher and A.A. Tseytlin, 
``Precision spectroscopy of AdS/CFT," JHEP \textbf{10} (2003) 037 
[hep-th/0308117].
\bibitem{AFRT} G. Arutyunov, S. Frolov, J. Russo and A.A. Tseytlin, 
``Spinning strings in $AdS_5 \times S^5$ and integrable systems,"
Nucl. Phys. \textbf{B671} (2003) 3 [hep-th/0307191].
\bibitem{ART} G. Arutyunov, J. Russo and A.A. Tseytlin, ``Spinning 
strings in $AdS_5\times S^5$: new integrable system relations," 
Phys. Rev. \textbf{D69} (2004) 086009 [hep-th/0311004].
\bibitem{AS} G. Arutyunov and M. Staudacher, ``Matching higher conserved
charges for strings and spins," JHEP \textbf{03} (2004) 004 
[hep-th/0310182]; ``Two-loop commuting charges and the string/gauge
duality," hep-th/0403077;
J. Engquist,``Higher conserved charges and integrability for
spinning strings in $AdS_5 \times S^5$," JHEP \textbf{04}
(2004) 002 [hep-th/0402092].
\bibitem{AT} A.A. Tseytlin, ``Spinning strings and AdS/CFT duality,"
hep-th/0311139.
\bibitem{SS} D. Serban and M. Staudacher, ``Planar $\mathcal{N} =4$ 
gauge theory and the Inozemtsev long range spin chain," JHEP \textbf{06}
(2004) 001 [hep-th/0401057].
\bibitem{MZ} J.A. Minahan and K. Zarembo, ``The Bethe-ansatz for 
$\mathcal{N} =4$ super Yang-Mills," JHEP \textbf{03} (2003) 013 
[hep-th/0212208];
 J. Engquist, J.A. Minahan and K. Zarembo, 
``Yang-Mills duals for semiclassical strings on $AdS_5 \times S^5$,''
 JHEP \textbf{11} (2003) 063 [hep-th/0310188];
J.A. Minahan, ``Higher loops beyond the SU(2) sector," 
JHEP \textbf{10} (2004) 053 [hep-th/0405243].
\bibitem{BKS} N. Beisert, C. Kristjansen and M. Staudacher, ``The 
dilatation operator of $\mathcal{N} =4$ super Yang-Mills theory," 
Nucl. Phys. \textbf{B664} (2003) 131 [hep-th/0303060].
\bibitem{NB} N. Beisert, ``The complete one-loop dilatation operator
of $\mathcal{N} =4$ super Yang-Mills theory," Nucl. Phys. \textbf{B676}
(2004) 3 [hep-th/0307015]; N. Beisert and M. Staudacher, ``The 
$\mathcal{N}=4$ SYM integrable super spin chain," Nucl. Phys. 
\textbf{B670} (2003) 439 [hep-th/0307042].
\bibitem{NBS} N. Beisert, ``Higher loops, integrability and the near
BMN limit," JHEP \textbf{09} (2003) 062 [hep-th/0308074];
``The su$(2|3)$ dynamic spin chain," Nucl. Phys. \textbf{B682}
(2004) 487 [hep-th/0310252].
\bibitem{BDS} N. Beisert, V. Dippel and M. Staudacher, ``A novel long
range spin chain and planar $\mathcal{N} =4$ super Yang-Mills,"
JHEP \textbf{07} (2004) 075 [hep-th/0405001];
G. Arutyunov, S. Frolov and M. Staudacher, ``Bethe
ansatz for quantum strings," JHEP \textbf{10} (2004)
016 [hep-th/0406256];
M. Staudacher, ``The factorized S-matrix of CFT/AdS," JHEP 
\textbf{05} (2005) 054 [hep-th/0412188].
\bibitem{NBP} N. Beisert, ``The dilatation operator of 
$\mathcal{N} = 4$ super Yang-Mills theory and integrability," 
Phys. Rept. \textbf{405} (2005) 1 [hep-th/0407277].
\bibitem{KMMZ} V.A. Kazakov, A. Marshakov, J.A. Minahan and K. Zarembo,
``Classical/quantum integrability in AdS/CFT," JHEP \textbf{05}
(2004) 024 [hep-th/0402207].
\bibitem{KZ} V.A. Kazakov and K. Zarembo, 
``Classical/quantum integrability in non-compact sector of AdS/CFT," 
JHEP \textbf{10} (2004) 060 [hep-th/0410105];
N. Beisert, V.A. Kazakov and K. Sakai, ``Algebraic curve for the SO(6)
sector of AdS/CFT," hep-th/0410253;
S. Sch\"afer-Nameki, ``The algebraic curve of 1-loop planar 
$\mathcal{N} = 4$ SYM," Nucl. Phys. \textbf{B714} (2005) 3
[hep-th/0412254];
N. Beisert, V.A. Kazakov, K. Sakai and K. Zarembo,
``The algebraic curve of classical superstrings on $AdS_5 \times S^5$,"
hep-th/0502226.
\bibitem{MK} M. Kruczenski, ``Spin chains and string theory,"
Phys. Rev. Lett. \textbf{93} (2004) 161602 [hep-th/0311203]. 
\bibitem{KRT} M. Kruczenski, A.V. Ryzhov and A.A. Tseytlin, ``Large
spin limit of $AdS_5 \times S^5$ string theory and low energy expansion
of ferromagnetic spin chains," Nucl. Phys. \textbf{B692} (2004) 3
[hep-th/0403120];
R. Hernandez and E. Lopez, ``The SU(3) spin chain sigma
model and string theory," JHEP \textbf{04} (2004) 052 [hep-th/0403139];
B. Stefanski, jr. and A.A. Tseytlin, ``Large spin limits
of AdS/CFT and generalized Landau-Lifshitz equations," 
JHEP \textbf{05} (2004) 042 [hep-th/0404133];
M. Kruczenski and A.A. Tseytlin, ``Semiclassical
relativistic strings in $S^5$ and long coherent operators in 
$\mathcal{N}=4$ SYM theory," JHEP \textbf{09} (2004) 038
[hep-th/0406189];
S. Bellucci, P.-Y. Casteill, J.F. Morales, C. Sochichiu,
``SL(2) spin chain and spinning strings on $AdS_5 \times S^5$,"
Nucl. Phys. \textbf{B707} (2005) 303 [hep-th/0409086];
S. Ryang ``Circular and folded multi-spin strings in spin chain
sigma models," JHEP \textbf{10} (2004) 059 [hep-th/0409217].
\bibitem{AAT} A.A. Tseytlin, ``Semiclassical strings and 
AdS/CFT,"  hep-th/0409296.
\bibitem{LM} O. Lunin and J. Maldacena, ``Deforming field theories
with $U(1)\times U(1)$ global symmetry and their gravity duals,"
JHEP \textbf{05} (2005) 033 [hep-th/0502086]. 
\bibitem{LS} R.G. Leigh and M.J. Strassler, ``Exactly marginal operators
and duality in four-dimensional $\mathcal{N}=1$ supersymmetric 
gauge theory," Nucl. Phys. \textbf{B447} (1995) 95 [hep-th/9503121]. 
\bibitem{NP} V. Niarchos and N. Prezas, ``BMN operators for 
$\mathcal{N}=1$ superconformal Yang-Mills theories and associated string
backgrounds," JHEP \textbf{06} (2003) 015 [hep-th/0212111]. 
\bibitem{SF} S. Frolov, ``Lax pair for strings in Lunin-Maldacena 
background," JHEP \textbf{05} (2005) 069 [hep-th/0503201].
\bibitem{AF} G. Arutyunov and S. Frolov, ``Integrable Hamiltonian
for classical strings on $AdS_5 \times S^5$," JHEP 
\textbf{02} (2005) 059 [hep-th/0411089];
L.F. Alday, G. Arutyunov and A.A. Tseytlin, ``On integrability of 
classical superstrings in $AdS_5 \times S^5$," JHEP \textbf{07} 
(2005) 002 [hep-th/0502240].
\bibitem{FRT} S.A. Frolov, R. Roiban and A.A. Tseytlin, ``Gauge-string
duality for superconformal deformation of $\mathcal{N}=4$ super 
Yang-Mills theory," hep-th/0503192.
\bibitem{RR} R. Roiban, ``On spin chains and field theories,"
JHEP \textbf{09} (2004) 023 [hep-th/0312218].
\bibitem{BC} D. Berenstein and S.A. Cherkis, ``Deformations of
$\mathcal{N}=4$ SYM and integrable spin chain models,"
Nucl. Phys. \textbf{B702} (2004) 49 [hep-th/0405215].
\bibitem{KMSS} R. de Mello Koch, J. Murugan, J. Smolic and 
M, Smolic, ``Deformed PP-waves from the Lunin-Maldacena background,"
JHEP \textbf{08} (2005) 072 [hep-th/0505227];
T. Mateos, ``Marginal deformation of $\mathcal{N}=4$ SYM and
Penrose limits with continuum spectrum," JHEP \textbf{08} (2005) 026
[hep-th/0505243].
\bibitem{BDR} N.P. Bobev, H. Dimov and R.C. Rashkov, ``Semiclassical
strings in Lunin-Maldacena background," hep-th/0506063.
\bibitem{BR} N. Beisert and R. Roiban, ``Beauty and the twist:
The Bethe ansatz for twisted $\mathcal{N}=4$ SYM,"
JHEP \textbf{08} (2005) 039 [hep-th/0505187];
S.A. Frolov, R. Roiban and A.A. Tseytlin, ``Gauge-string
duality for (non)supersymmetric deformations of $\mathcal{N}=4$ super 
Yang-Mills theory," JHEP \textbf{07} (2005) 045 [hep-th/0507021];
J.G. Russo, ``String spectrum of curved string backgrounds obtained
by T-duality and shifts of polar angles," hep-th/0508125;
R. de Mello Koch, N. Ives, J. Smolic and M. Smolic, ``Unstable giants,"
hep-th/0509007;
M. Spradlin, T. Takayanagi and A. Volovich, ``String theory in 
$\beta$ deformed spacetimes," hep-th/0509036;
R.C. Rashkov, K.S. Viswanathan and Y. Yang, ``Ts...sT transformation
on $AdS_5 \times S^5$ background," hep-th/0509058;
J.A. Minahan, A. Tirziu and A.A. Tseytlin, ``1/J corrections to 
semiclassical AdS/CFT states from quantum Landau-Lifshitz model,"
hep-th/0509071.
\bibitem{FG} P.Z. Freedman and U. Gursoy, ``Comments on the 
$\beta$-deformed $\mathcal{N}=4$ SYM theory," hep-th/0506128;
S. Penati, A. Santambrogio and D. Zanon, ``Two-point correlators in
the $\beta$-deformed $\mathcal{N}=4$ SYM at the next-to-leading order,"
hep-th/0506150;
G. Rossi, E. Sokatchev and Y.S. Stanev, ``New results in the deformed
$\mathcal{N}=4$ SYM theory," hep-th/0507113;
A. Mauri, S. Penati, A. Santambrogio and D. Zanon, ``Exact results in
planar $\mathcal{N}=1$ superconformal Yang-Mills theory," hep-th/0507282;
S.M. Kuzenko and A.A. Tseytlin, ``Effective action of $\beta$-deformed
$\mathcal{N}=4$ SYM theory and AdS/CFT," hep-th/0508098.
\bibitem{GN} U. Gursoy and Nunez, ``Dipole deformations of 
$\mathcal{N}=1$ SYM and supergravity backgrounds with $U(1) \times U(1)$
global symmetry," Nucl. Phys. \textbf{B725} (2005) 45 [hep-th/0505100];
C. Ahn and J.F. Vazquez-Poritz, ``Deformations of flows from type IIB
supergravity," hep-th/0508075.



\end{thebibliography}
\end{document}